\documentclass[pra,onecolumn,floatfix,a4paper,superscriptaddress]{revtex4}
\usepackage{bm,color,graphicx,amsmath,txfonts}

\usepackage[colorlinks, citecolor=blue,linkcolor=blue]{hyperref}

\newcommand{\Tr}{{\rm Tr}}

\newcommand{\e}{{e}}

\begin{document}

\title{Detection of entanglement by harnessing extracted work in an opto-magno-mechanics}

\author{M'bark Amghar}     
\affiliation{Department of Physics, Ibnou Zohr University, Agadir 80000, Morocco}

\author{Mohamed Amazioug}     \thanks{amazioug@gmail.com}
\affiliation{Department of Physics, Ibnou Zohr University, Agadir 80000, Morocco}

\begin{abstract}

The connections between thermodynamics and quantum information processing are of paramount importance. Here, we address a bipartite entanglement via extracted work in a cavity magnomechanical system contained inside an yttrium iron garnet (YIG) sphere. The photons and magnons interact through an interaction between magnetic dipoles. A magnetostrictive interaction, analogous to radiation pressure, couple’s phonons and magnons. The extracted work was obtained through a device similar to the Szil\'ard engine. This engine operates by manipulating the photon-magnon as a bipartite quantum state. We employ logarithmic negativity to measure the amount of entanglement between photon and magnon modes in steady and dynamical states. We explore the extracted work, separable work, and maximum work for squeezed thermal states. We investigate the amount of work extracted from a bipartite quantum state, which can potentially determine the degree of entanglement present in that state. Numerical studies show that entanglement, as detected by the extracted work and quantified by logarithmic negativity, is in good agreement. We show the reduction of extracted work by a second measurement compared to a single measurement. Also, the efficiency of the Szilard engine in steady and dynamical states is investigated. We hope this work is of paramount importance in quantum information processing.

\end{abstract}

\date{\today}
\maketitle

\section{Introduction}

Entanglement, a cornerstone of quantum mechanics pioneered in works like \cite{EPR35,Schrodinger35,Bell64}, holds immense potential across various fields. It has exciting applications in areas like precision measurement, quantum key distribution \cite{QKD91}, teleporting quantum information \cite{Bennett93}, and building powerful quantum computers as explored by \cite{Bennett00}. Entanglement creates a spooky connection between particles. Measuring one instantly affects the other, defying classical ideas of locality.

Cavity optomechanics is a field of physics that studies the interaction between light and mechanical objects via radiation pressure at very small scales~\cite{OMRMP}. Cavity optomechanics has been used to develop new methods for generating and manipulating squeezed states of light, which are a type of quantum state. Recently this cavity has paramount importance applications in quantum information processing such as quantum entangled states \cite{DVitali2007,Abdi1,Hichem14,Abdi2,sete14,asjad2018, MAmaziougEPJD2018, asjad2013,MAmaziougQIP2020,Berihu18,Tesfay,Habtamu}, cooling the mechanical mode to their quantum ground states \cite{JTeufel2011, SMachnes2012, asjadC19, JChan2011, MBhattacharya2007}, photon blockade \cite{MAmaziougPB2022}, enhancing precision measurements \cite{HXiong17,Candeloro}, superconducting elements~\cite{supercond1}, and also between two massive mechanical oscillators~\cite{enMM1,enMM2} have been observed. Cavity quantum electrodynamics (CQED) paved the way for the emergence of a new field called cavity optomechanics. Cavity quantum electrodynamics (CQED) offers control over how light (photons) interacts with atoms at the quantum level. Actually, single quanta can significantly impact the atom-cavity dynamics in the strong coupling regime, which is made possible by strong confinement. Recently, we have successfully achieved strong coupling in numerous experiments, leading to the demonstration of fascinating quantum phenomena such as quantum phase gates \cite{QATurchette95}, the Fock state generation \cite{BTHVarcoe00}, and quantum nondemolition detection of a single cavity photon \cite{GNogues99}. Building on the success of cavity QED, exploring how magnon systems interact within cavity optomechanics offers a promising avenue for unlocking their unique quantum properties. The first experimental demonstration of interaction between magnons, photons, and phonons has been achieved~\cite{Tang16}. This system combines magnon-photon coupling, similar to what's found in magnon QED, with an additional coupling between magnons and phonons. While the cavity output reflects the impact of magnon-phonon coupling, a more comprehensive understanding would require a full quantum treatment that incorporates these fluctuations. From the standpoint of cavity quantum electrodynamics (QED), ferrimagnetic systems. Particularly, the yttrium iron garnet (YIG) sphere has garnered a lot of attention. Studies have shown that the YIG sphere's Kittel mode~\cite{Kittel} can achieve strong coupling with microwave photons trapped within a high-quality cavity. This strong coupling leads to the formation of cavity polaritons~\cite{Strong1} and a phenomenon known as vacuum Rabi splitting. The success of cavity QED has opened doors to applying many of its concepts to the emerging field of magnon cavity QED~\cite{Nori}. This new field has already seen exciting advancements, including the observation of bistability~\cite{You18} and the groundbreaking coupling of a single superconducting qubit to the Kittel mode~\cite{Tabuchi}. Recently, magnons (as spin waves) have been studied extensively in the field of quantum information processing \cite{Hussain22,asjad23,amaziougsr,amaziougpy}.

Studies suggest that physical system information can be utilized to extract work with suitable operations \cite{Maruyama09}. This, described as information thermodynamics, explores the relationship between information theory and thermodynamics. By studying how information can be manipulated to perform physical tasks, researchers hope to uncover new ways to improve efficiency in various processes. Szilard's engine \cite{Parrondo15} is a classic example of this concept. It demonstrates how information processing is capable of extracting work from a physical system. Interestingly, research has shown that the specific way information is encoded, particularly in entangled or correlated states, plays a crucial role in how much work can be extracted \cite{Oppenheim02,Maruyama05,Funo13}.

In this letter, we investigate the potential of exploiting both extracted work and efficiency in optomagnomechanical systems. This study reveals the presence of entanglement between magnons and cavity photons in an optomechanical system. We achieve this detection by examining the extractable work. Our work breaks new ground by employing extractable work as a tool to detect entanglement between magnons and cavity photons in optomechanical systems. Our study, utilizing realistic experimental parameters, reveals excellent agreement between the entanglement region detected via extractable work and the results of Jie Li et al \cite{JLi18}. This highlights the validity of our method for entanglement detection in magnomechanical systems. This applies not only to steady-state conditions but also to dynamical states under thermal influence. Our study further explores the influence of various parameters, including detuning and magnon-phonon coupling, on the entanglement properties of the system. In addition, we analyze the information-work efficiency under thermal noise, considering both steady-state and dynamical regimes.

The article is outlined as follows: In Section II, we introduce the model for the optomagnomechanical system, its Hamiltonian, and the quantum nonlinear Langevin equations for the interacting photon-magnon-phonon system. Section III tackles the linearization of the QLEs, and we then assess the covariance matrix for steady and dynamical states. Section IV delves into the connection between quantum thermodynamics and quantum entanglement in cavity magnomechanical systems. We employ logarithmic negativity to quantify the entanglement between the photon and magnon modes and investigate how it relates to the amount of work that can be extracted from the system. Besides, we have also investigated the efficiency of a Szilard engine. The results obtained are discussed. Concluding remarks close this paper.


\section{Model}

In this section, we consider on a system that combines a microwave cavity with magnetic excitations (magnons) and mechanical vibrations (phonons). This cavity magnomechanical system is illustrated in Fig.~\ref{fig1}. Magnons are a collective motion of numerous spins in a ferrimagnet, such as an YIG sphere (250-$\mu$m-diameter sphere used in Ref.~\cite{Tang16}). A sphere made of YIG (Yttrium Iron Garnet) is positioned within a microwave cavity at a location with the strongest magnetic field. Additionally, a uniform bias magnetic field is applied throughout the entire system. These combined fields allow the microwave photons in the cavity to interact with the YIG sphere's magnons through the magnetic dipole interaction. To improve the coupling between magnons and phonons, the experiment utilizes a microwave source (not shown) to directly driven the magnon mode magnetostrictive interaction. Due to the YIG sphere's small size compared to the microwave wavelength, we can ignore the interaction between microwave photons and phonons. We consider the three magnetic fields: a bias field pointing in the z-axis, a drive field in the y-axis, and the magnetic field of the cavity mode oriented along the x-axis, as depicted in Fig.~\ref{fig1}. These three fields are mutually perpendicular at the position of the YIG sphere. The YIG sphere experiences a deformation of its geometry structure due to the creation of vibrational modes, or phonons, which influence the magnon excitations within the sphere, and vice versa~\cite{Kittel2}. The Hamiltonian writes as \cite{JLi18}
\begin{equation}
\begin{split}
 \mathcal{\hat H} &= \hbar\Omega_c \hat c^{\dag} \hat c + \hbar\Omega_n \hat n^{\dag} \hat n + \frac{\Omega_d}{2} (\hat x^2 + \hat y^2)\\
& + \hbar g_{nd} \hat n^{\dag} \hat n \hat x + \hbar \mathcal{G}_{nc} (\hat c + \hat c^{\dag}) (\hat n + \hat n^{\dag})\\
& + i \hbar\Omega (\hat n^{\dag} e^{-i \Omega_0 t}  - \hat n e^{i \Omega_0 t} ),
\end{split}
\end{equation}
where the creation and annihilation operators for the cavity ($\hat c, \hat c^{\dag}$) and magnon ($\hat n, \hat n^{\dag}$) modes satisfy the canonical commutation relation $[\mathbb{\hat O}, \mathbb{\hat O}^{\dag}]\,{=}\,1$ (where $\mathbb{\hat O}$ can be $\hat c$ or $\hat n$).  Additionally, dimensionless position $\hat x$ and momentum $\hat y$ operators for the mechanical mode are included, with the commutation relation $[\hat x, \hat y]\,{=}\,i$. The Hamiltonian also incorporates the resonance frequencies ($ \Omega_c$, $\Omega_n$, and $\Omega_d$) of the cavity, magnon, and mechanical modes, respectively. The magnon frequency $\Omega_n$ is dictated by the external bias magnetic field $H$ and the gyromagnetic ratio $\gamma$ following the relation: $\Omega_n=\gamma H$.  Interestingly, the magnon-microwave coupling rate $\mathcal{G}_{nc}$ surpasses the dissipation rates of both the cavity $\lambda_c$ and magnon modes $\lambda_n$, satisfying the condition for strong coupling: $\mathcal{G}_{nc} > \lambda_c, \lambda_n$~\cite{Strong1}. The inherent coupling rate between a single magnon and the mechanical vibrations, denoted by $g_{nd}$, is typically low.  This limitation can be overcome by strategically applying a strong microwave field directly to the YIG sphere. This approach, employed in earlier works~\cite{You18,You16}, effectively enhances the magnomechanical interaction. The Rabi frequency $\Omega$, derived under the assumption of low-lying excitations ($\langle \hat n^{\dag} \hat n \rangle \ll 2\mathcal{N} s$, where $s=5/2$ is the spin of the Fe$^{3+}$ ground state ion), characterizes the coupling strength between the driving magnetic field (amplitude $B_0$ and frequency $\omega_0$) and the magnon mode. It is expressed as $\Omega =\frac{\sqrt{5}}{4} \gamma_g \! \sqrt{\mathcal{N}} B_0$, where $\gamma_g/2\pi= 28$ GHz/T is the gyromagnetic ratio of the material and $\mathcal{N}=\varrho \mathcal{V}$ represents the total number of spins in the YIG sphere. Here, $\varrho=4.22 \times 10^{27}$ m$^{-3}$ is the spin density and $\mathcal{V}$ is the sphere's volume. Using a rotating frame at the driving frequency $\Omega_0$ and the rotating-wave approximation $\mathcal{G}_{nc} (\hat c + \hat c^{\dag}) (\hat n + \hat n^{\dag}) \to \mathcal{G}_{nc} (\hat c \hat n^{\dag} + \hat c^{\dag} \hat n)$ valid when $\Omega_c, \Omega_n \gg \mathcal{G}_{nc}, \kappa_{c}, \kappa_{n}$~\cite{Tang16}, the system's dynamics are described by quantum Langevin equations (QLEs). 
\begin{equation}
\begin{split}
\delta\dot{\hat c}&= - (i \delta_c + \lambda_c) \hat c - i \mathcal{G}_{nc} \hat n + \sqrt{2 \lambda_c} \hat c^{\rm in},  \\
\delta\dot{\hat n}&= - (i \delta_n + \lambda_n) \hat n - i \mathcal{G}_{nc} \hat c - i g_{nd} \hat n \hat x + \Omega + \sqrt{2 \lambda_n} \hat n^{\rm in},  \\
\delta\dot{\hat x}&= \omega \hat y, \quad \delta\dot{\hat y}= - \omega \hat x - \gamma_d \hat y - g_{nd} \hat n^{\dag}\hat n + \eta, 
\end{split}
\end{equation}
with $\delta_{c}=\Omega_{c}-\Omega_0$, $\delta_{n}=\Omega_{n}-\Omega_0$ and $\gamma_d$ is the mechanical damping rate. The input noise operators for the cavity and magnon modes are respectively, $\hat c^{\rm in}$ and $\hat n^{\rm in}$, with zero mean, i.e., $\langle \hat c^{\rm in} \rangle=\langle \hat n^{\rm in} \rangle=0$, and described by the following correlation functions \cite{Zoller}:
\begin{equation} \label{noises}
\begin{split}
\langle \hat c^{\rm in}(t) \, \hat c^{\rm in \dag}(t')\rangle = [N_c(\Omega_c){+}1] \,\delta(t{-}t')\quad,\quad \langle \hat c^{\rm in \dag}(t) \, \hat c^{\rm in}(t')\rangle \,\,{=}\,\, N_c(\Omega_c) \, \delta(t{-}t')  \\
\langle \hat n^{\rm in}(t) \, \hat n^{\rm in \dag}(t')\rangle = [N_n(\Omega_n)+1] \, \delta(t{-}t')\quad,\quad\langle \hat n^{\rm in \dag}(t) \, \hat n^{\rm in}(t')\rangle = N_n(\Omega_n)\, \delta(t{-}t'),
\end{split}
\end{equation}
we assume the mechanical mode follows a Markovian process. This means a large mechanical quality factor $Q \,\, {\gg}\, 1$, i.e., $\Omega_d \,\, {\gg}\, \gamma_d$~\cite{Markov}. Furthermore, the noise operators for this mode possess non-zero correlation properties (with $\langle \eta(t) \rangle=0$), writes as
\begin{equation} \label{noises}
\langle \eta(t)\eta(t')\,{+}\,\eta(t') \eta(t) \rangle/2 \,\, {\simeq} \,\, \gamma_d [2 N_d(\Omega_d) {+}1] \delta(t{-}t'),
\end{equation}
where $N_c$, $N_n$ and $N_d$ correspond to the equilibrium mean thermal occupation numbers for the cavity photons, magnons, and phonons, respectively. Thus $N_j(\Omega_j){=}\big[ {\rm exp}\big( \frac{\hbar \Omega_j}{k_B T} \big) {-}1 \big]^{-1} $ $(j{=}\hat c, \hat n, \hat d)$, where $k_B$ is the Boltzmann constant.

\section{Covariance matrix}

We consider the case where the magnon mode is highly driven. We linearize the non-linear quantum Langevin equation by assuming small fluctuations around a steady state amplitude, i.e., $\mathbf{\hat O} = \mathbf{\hat O}_{ss} +\delta \mathbf{\hat O}$ ($\mathbf{\hat O}\, {=}\, \hat c, \hat n, \hat x, \hat y$), where $\hat n_{ss}$ writes as
\begin{equation}\label{eq5}
\hat n_{ss} =  \frac{\Omega  ( i \delta_c +\lambda_c) }{\mathcal{G}_{nc}^2 \! + ( i \tilde{ \delta}_n + \lambda_n) ( i \delta_c + \lambda_c) },
\end{equation}
where $\bar{\delta}_n = \delta_n + g_{nd} \hat x_{ss}$ is the effective magnon-drive detuning taking into account a frequency shift resulting from the interaction between magnons and phonons. This interaction is known as magnomechanical interaction. Under the condition $|\bar{ \delta}_n|, |\delta_c| \gg  \lambda_c, \lambda_n$; $\hat n_{ss}$ is given by
\begin{equation}\label{nss}
\hat n_{ss} =  \frac{ i  \Omega  \delta_c} {\mathcal{G}_{nc}^2  - \tilde{ \delta}_n \delta_c }.
\end{equation}
and $\hat x_{ss} = - \frac{g_{nd}}{\Omega_d} |\hat n_{ss}|^2 $. The system is described by linearized quantum Langevin equations (LQLEs)
\begin{equation} \label{LQLEs}
\begin{split}
\delta\dot{\hat c}&= - (i \delta_c + \lambda_c) \hat c - i \mathcal{G}_{nc} \hat n + \sqrt{2 \lambda_c} \hat c^{\rm in},  \\
\delta\dot{\hat n}&= - (i \bar\delta_n + \lambda_n) \hat n - i \mathcal{G}_{nc} \hat c - i g_{nd} \hat n_{ss} \hat x + \omega + \sqrt{2 \lambda_n} \hat n^{\rm in},  \\
\delta\dot{\hat x}&= \omega \hat y, \quad \delta\dot{\hat y}= - \omega \hat x - \gamma_d \hat y - g_{nd} (\hat n_{ss} \hat n^{\dag}+\hat n^*_{ss}\hat n) + \eta, 
\end{split}
\end{equation}
where $\hat n_{ss}=- \frac{i\mathcal{G}_{nd}}{\sqrt{2}g_{nd}}$ is the magnon-phonon coupling. The quadrature fluctuations $(\delta X_c, \delta Y_c, \delta X_n, \delta Y_n, \delta x, \delta y)$ are described as 
$$\delta X_c=(\delta \hat c + \delta \hat c^{\dag})/\sqrt{2}\quad,\quad \delta Y_c=i(\delta \hat c^{\dag} - \delta \hat c)/\sqrt{2}$$
$$\delta X_n=(\delta \hat n + \delta \hat n^{\dag})/\sqrt{2}\quad,\quad \delta Y_n=i(\delta \hat n^{\dag} - \delta \hat n)/\sqrt{2}$$
We can rewrite equation (\ref{LQLEs}) as
\begin{equation}
\dot{v} (t) = \mathcal{F} v(t) + \chi(t) ,
\end{equation}
where $v(t)=\big[\delta X_c (t), \delta Y_c (t), \delta X_n (t), \delta Y_n (t), \delta x (t), \delta y (t) \big]^T$ is the quadrature vector, $\chi (t) = \big[ \!\sqrt{2\lambda_c} X_c^{\rm in} (t), \sqrt{2\lambda_c} Y_c^{\rm in} (t), \sqrt{2\lambda_n} X_n^{\rm in} (t), \sqrt{2\lambda_n} Y_n^{\rm in} (t), 0, \eta (t) \big]^T$ is the noise vector and the drift matrix $\mathcal{F}$ is written as 
\begin{equation}\label{AAA}
\mathcal{F} =
\begin{pmatrix}
-\lambda_c  &  \delta_c  &  0 &  \mathcal{G}_{nc}  &  0  &  0   \\
-\delta_c  & -\lambda_c  & -\mathcal{G}_{nc}  & 0  &  0  &  0   \\
0 & \mathcal{G}_{nc}  & -\lambda_n  & \tilde{ \delta}_n &  -\mathcal{G}_{nd}  &  0 \\
-\mathcal{G}_{nc}  & 0 & -\tilde{ \delta}_n & -\lambda_n &  0  &  0 \\
0 &  0  &  0  &  0  &  0  &  \omega_d   \\
0 &  0  &  0  &  \mathcal{G}_{nd}  & -\omega_d & -\lambda_d   \\
\end{pmatrix} ,
\end{equation}
The system under consideration is considered stable if all the eigenvalues of a drift matrix, have real parts that are negative~\cite{stability87}. By using the Lyapunov equation, the system's state is expressed as~\cite{DV07,Hahn}
\begin{equation}\label{Lyap}
\mathcal{F} \mathcal{C}+\mathcal{C} \mathcal{F}^T = -\mathcal{L},
\end{equation}
where $\mathcal{C}_{ij}=\frac{1}{2}\langle v_i(t) v_j(t') + v_j(t') v_i(t)   \rangle$ ($i,j=1,2,...,6$) is the covariance matrix and $\mathcal{L}={\rm diag} \big[ \lambda_c (2N_c+1), \lambda_c (2N_c+1), \lambda_n (2N_n+1),  \lambda_n (2N_n+1), 0,  \lambda_d (2N_d +1 ) \big]$ is the diffusion matrix achieved through $\mathcal{L}_{ij} \delta (t-t')=\frac{1}{2}\langle \chi_i(t) \chi_j(t') + \chi_j(t') \chi_i(t) \rangle $.

\section{Szil\'ard engine}    \label{s:Szilard}

L\'eo Szil\'ard introduced Szil\'ard's engine as a thought experiment in 1929. This experiment simplified the famous Maxwell's demon paradox by using just one molecule of gas and replacing the demon with a mechanical device. Szil\'ard's engine operates in four key steps: (i) The experiment begins with a single gas molecule bouncing around freely in a container with a volume of $V$. (ii) A separator is placed inside a container, dividing it into two equal chambers with a volume of $V/2$, ensuring no heat exchange during the process. (iii) The engine's function relies on determining the molecule's location in the left or right chamber. According to the measurement results, a tiny weight has been attached to the same side of the partition using a pulley system. (iv) The final stage involves connecting the entire setup to a constant temperature heat source, allowing a gas molecule to expand and fill the container, crucial for the engine's theoretical work. Szil\'ard's engine, a concept that challenges our understanding of thermodynamics, involves a single molecule expanding to fill a container, absorbing heat from a constant temperature bath, and converting this heat into usable work by lifting the weight attached to the partition. The amount of work extracted, can be calculated using the formula $W=k_B T \ln 2$, where $k_B$ is Boltzmann's constant and $ln(2)$ represents the information gain from measuring the molecule's location. This process relies on connecting a weight to the partition and allowing the single gas molecule to expand in a controlled, constant temperature, i.e., isothermal manner. Szil\'ard's engine can theoretically extract a specific amount of work per cycle, as described in~\cite{Plenio}
\begin{equation}\label{S}
W = k_B T \ln 2 [1-H(X)]
\end{equation}
The uncertainty about where the molecule is situated (left or right) can be quantified using a concept called Shannon entropy. This entropy is denoted by $H(X) = -\sum_x p_x \ln p_x$, where $p_x$ is the probability of capturing the molecule in each location ($x=R$ or $x=L$). Thus, the more uncertain we are about the molecule's location, the higher the Shannon entropy will be. Equation (\ref{S}) presents a potential challenge to the second law of thermodynamics. It suggests that under specific conditions, perfect knowledge about a the microscopic information of the system state might allow for work extraction using only a single heat bath. A significant link between information processing and the physical world was suggested by physicist Rolf Landauer in 1961. He theorized that whenever a single bit of information is erased in a computer system, it leads to an increase in energy dissipation as heat. This principle suggests a fundamental link between logical operations within a computer and the laws of thermodynamics that govern physical processes \cite{Landauer}. Recent experiments explore innovative techniques inspired by Maxwell's demon and Landauer's principle, which link information processing and energy dissipation, despite the potential to defy thermodynamic laws~\cite{Exp15,Exp16}. By separating two entangled particles into different containers, we essentially create two Szil\'ard engines, $A$ and $B$. These engines are unique because their functionality is intrinsically linked. Thus, the amount of work extractable from engine $A$ is dependent on the specific state of its entangled partner in engine $B$ 
\begin{equation}\label{S2}
W(A\vert B)=k_B T\log[1-H(A\vert B)]
\end{equation}

Due to the entanglement, any event affecting engine $B$ has an immediate impact on our understanding of engine $A$. Mutual information $I(A:B)=H(A)-H(A\vert B)\ge 0$ quantifies the link between $A$ and $B$, with a non-negative value indicating that knowing the state of $B$ reduces uncertainty about $A$. The reduced uncertainty leads to a significant increase in work extraction from engine $A$ compared to a scenario without entangled two Szil\'ard engines, as indicated by $W(A\vert B)$ being greater than or equal to $W(A)$, i.e., $W(A\vert B)\ge W(A)$.

\section{Negativity logarithmic, Work extraction and efficiency}

In this section, we will quantify and harness negativity logarithmic compared to the extracted work in a two-mode Gaussian state, shared by Alice ($A$ : photon) and Bob ($B$ : magnon). The efficiency of a Szilard engine will be adopted. 

\subsection{Negativity logarithmic}

The covariance matrix corresponding to the photon and magnon modes in the $(\delta X_c (t), \delta Y_c (t), \delta X_n (t), \delta Y_n (t))$ basis can be expressed as
\begin{equation}\label{VAB}
\mathcal{C}_{AB}=
\left(
\begin{array}{cc}
\mathcal{C}_c \, & \mathcal{C}_{cn} \\[1ex]
\mathcal{C}_{cn}^T &  \mathcal{C}_n
\end{array}
\right) \, ,
\end{equation}
$\mathcal{C}_{c}$ and $\mathcal{C}_{n}$ depict the covariance matrix $2\times 2$, respectively, representing the photon mode and magnon mode. The correlations between photon and magnon modes in standard form are denoted by $\mathcal{C}_{cn}$
\begin{equation}\label{Csf}
\mathcal{C}_c=\text{diag}(\alpha,\alpha) \,, \quad \mathcal{C}_n=\text{diag}(\beta,\beta) \,, \quad \mathcal{C}_{cn}=\text{diag}(\Delta,-\Delta) \,.
\end{equation}
For measuring bipartite entanglement, we employ the logarithmic negativity $E_N$~\cite{Plenio,GVidal02,GAdesso04}, that is given by 
\begin{equation} \label{eq:37}
E_{om} = \max[0,-\log(2\nu^-)],
\end{equation}
where $\nu^-= \sqrt{\mathcal{Y}-(\mathcal{Y}^2-4\det\mathcal{C}_{AB})^{1/2}}/\sqrt{2}$ is the minimum symplectic eigenvalue of the $\mathcal{C}_{AB}$, where $\mathcal{Y}=\det \mathcal{C}_c+\det \mathcal{C}_n-2\det \mathcal{C}_{cn}$.

\subsection{Magnon only performs Gaussian measurement}

The medium under consideration is a two-mode Gaussian state, i.e., photon and magnon modes. When Bob executes a Gaussian measurement on his assigned part of the system, the measurement has an impact on Alice's state. This measurement can be described by
\begin{equation}\label{MG}
\tilde N_n(X)=\pi^{-1}\tilde D_n (X)\tilde{\rho}^{N_n}\tilde D_n^{\dagger}(X),
\end{equation} 
where $\tilde D_n(X)=\e^{X\delta\hat{n}^{\dagger}-X^* \delta\hat{n}}$ is the displacement operator, $\tilde{\rho}^{N_n}$ is a pure Gaussian state without first moment and the its covariance matrix is given by
\begin{equation}\label{MGcv}
\Gamma^{N_n}=\frac{1}{2}R(\xi)\text{diag}(\lambda,\lambda^{-1})R(\xi)^T,
\end{equation} 
where $\lambda$ is a positive real number, $R(\xi)=[\cos\xi,-\sin\xi;\sin\xi,\cos\xi]$ is a rotation matrix and the detection of homodyne (heterodyne) is suggested by $\lambda = 0~~(\lambda = 1)$, individually. The outcome $X$ Bob gets from his measurement, it doesn't affect the state of Alice's mode $\delta\hat{c}$, i.e., $\mathcal{C}^{N_n}_{c\vert X}=\mathcal{C}^{N_n}_{c}$. The constrained state of mode $A$'s covariance matrix can be explicitly expressed as
\begin{equation}\label{CNn}
\mathcal{C}_c^{N_n}=\mathcal{C}_c- \mathcal{C}_{cn}(\mathcal{C}_n+\Gamma^{N_n})^{-1}\mathcal{C}_{cn}^T \, .
\end{equation} 
Bob measurement does push the state of mode $A$ out of equilibrium. However, by interacting with a heat bath for long time, mode $A$ eventually returns to an equilibrium state $\mathcal{C}_c^{\text{eq}}$. Its average entropy is solely $\int\text{d}X p_X S(\mathcal{C}_{c\vert X}^{N_n})=S(\mathcal{C}_c^{N_n})$ because her state is unaffected by the result. Work can be extracted by Alice from a surrounding heat bath~\cite{Mauro17}
\begin{equation}\label{WorkExt}
W=k_B T \left[ S(\mathcal{C}_c^{\text{eq}})-S(\mathcal{C}_c^{N_n})\right].
\end{equation}
We adopt the case of the covariance matrix in a squeezed thermal state, as depicted in equation (\ref{Csf}) and $\mathcal{C}_c^{\text{eq}}=\mathcal{C}_c$. The entropy of the covariance matrix described by equation \eqref{CNn} is quantified by considering the second-order R\'enyi entropy $S_2(\varrho)=-\ln Tr{\varrho^2}$ \cite{Adesso12}. In the case of two modes, Gaussian states (see equation (\ref{Csf})) are written as
\begin{equation}\label{ShanWig}
S_2(\mathcal{C}_{AB})=\frac12 \ln (\det \mathcal{C}_{AB}) \, .
\end{equation}
The extracted work, Eq.~\eqref{WorkExt}, became
\begin{equation}\label{WorkExtRenyi}
W^{(\lambda)}=\frac{k_B T}{2}\ln \left(\frac{\det \mathcal{C}_c}{\det \mathcal{C}_c^{N_n}}\right) \, .
\end{equation}
The extractable work for both homodyne ($\lambda=0$) and heterodyne ($\lambda=1$) detection in the case of STSs, writes as
\begin{equation}\label{ShanWig}
W_{om}^{(0)}=\frac{k_B T}{2} \ln \left(\frac{\alpha\beta}{\alpha\beta-\Delta^2}\right)\quad,\quad W_{\text{omSep}}^{(0)}=\frac{k_B T}{2} \ln \left(\frac{4\alpha\beta}{2\alpha+2\beta-1}\right)\quad,\quad W_{\text{omMax}}^{(0)}=\frac{k_B T}{2} \ln \left[\frac{4\alpha\beta}{1+2\vert \alpha-\beta\vert}\right].
\end{equation}

\begin{equation}\label{WorkSTS}
W_{om}^{(1)} = k_B T \ln \left[\frac{2\alpha\beta+\alpha}{2\alpha\beta+\alpha-2\Delta^2}\right]\quad,\quad W_{\text{omSep}}^{(1)}=k_B T \ln \left[\frac{2\alpha(2\beta+1)}{4\alpha+2\beta-1}\right] \quad,\quad W_{omMax}^{(0)}= 
\begin{cases}
    k_B T\ln 2\alpha  & \mathrm{if} \; \; \alpha \le \beta\\
    k_B T\ln\left[\frac{2\alpha(1+2\beta)}{1+4\alpha-2\beta} \right] & \mathrm{otherwise}
\end{cases}
\end{equation} 
The works remain independent of the measurement angle.

\subsection{Both magnon and photon perform Gaussian measurement}

This subsection explores the case where Alice and Bob, each make Gaussian measurements on their state. Alice now performs a second Gaussian measurement on her reduced state of the system, it can be described by
\begin{equation}\label{MG}
\tilde N_c(X)=\pi^{-1}\tilde D_c (Y)\tilde{\rho}^{N_c}\tilde D_c^{\dagger}(Y),
\end{equation} 
where $\tilde D_c(Y)=\e^{Y\delta\hat{c}^{\dagger}-Y^* \delta\hat{c}}$ is the displacement operator, $\tilde{\rho}^{N_c}$ corresponds to a pure Gaussian state without first moment and the its covariance matrix is given by
\begin{equation}\label{MGcv}
\Gamma^{N_n}=\frac{1}{2}R(\chi)\text{diag}(\Lambda,\Lambda^{-1})R(\chi)^T,
\end{equation} 
where $R(\chi)$ represents a rotation matrix and $\Lambda \in [0,\infty]$. The probability distribution describing a Gaussian measurement on Alice mode $\delta\hat{c}$ is influenced by the measurement $\tilde N_n(X)$ performed on Bob mode $\delta\hat{n}$. However, interestingly, the uncertainty in Alice mode $\delta\hat{c}$ remains unaffected by the outcome $X$ that Bob obtains from his Gaussian measurement, i.e., $\mathcal{C}_{cn}^{N_n, N_c}=\mathcal{C}_{c}^{N_n}+\Gamma^{N_c}$, While $\mathcal{C}_{c}^{N_n}$ is provided by Eq. (\ref{CNn}). The extracted work by Alice (photon), can be measured via the Shannon entropy of the appropriate probability distribution $H(\text{Pr}(X,Y))$ is similar to the entropy of the Gaussian distribution $H(\mathcal{C}_{cn}^{N_n, N_c})$. Its expression writes as
\begin{align}\label{WorkTwoMeas}
W^{(\lambda,\Lambda)}(\xi,\chi)=\frac{k_B T}{2}\ln \left(\frac{\det \mathcal{C}_c^{N_n}}{\det\mathcal{C}_{cn}^{N_n, N_c}}\right) \, .
\end{align}
In the case of STSs the extractable work for both homodyne ($\lambda=0$) and heterodyne ($\lambda=1$), writes as
\begin{align}\label{WorkTwoMeas}
W_{om}^{(0,0)}(\xi,\chi)&=k_B T\ln \left[\sqrt{\frac{4\alpha\beta}{4\alpha\beta-2\Delta^2[1+\cos(2\xi+2\chi)]}}\right] \, , \quad W_{om}^{(1,1)}=k_B T\ln \left[\frac{(1+2\alpha)(1+2\beta)}{1+2\beta+\alpha(2+4\beta)-4\Delta^2}\right] \, .
\end{align}

\subsection{Efficiency of the work extraction}

According to Zhuang et al. (2014), the information-work efficiency of a Szilard engine can be expressed as the ratio of extracted work to erasure work \cite{Zhuang14}
\begin{equation}\label{efficiency1}
\mu = \frac{W}{W_{eras}},
\end{equation} 
In this case, the information contained in the system is proportionate to $W_{eras}$ 
\begin{equation}\label{efficiency2}
W_{eras} = k_B T H(P)\ln 2,
\end{equation} 
where the Shannon entropy connected to the probability $P_j$ distribution is expressed as ??
We exploit the density operators $\rho$ and the von Neumann entropy to serve as the quantum mechanical equivalents of probability distributions \cite{Marina23}
\begin{equation}\label{efficiency3}
S(\rho) = \Tr(\rho\log_ (\rho)),
\end{equation} 
For two modes Gaussian state $\rho_G$ the von Neumann entropy $s_V$ can be written as 
\begin{equation}\label{efficiency3}
S(\rho_G) = \sum^2_{l=1} s_V(\Phi_l),
\end{equation} 
with $\Phi_l$, $l = 1,2$, represent the symplectic eigenvalues of the matrix $\mathcal{C}_{AB}$ (see equation \eqref{VAB}) writes as
\begin{equation}\label{efficiency3}
\Phi^{\pm} = \sqrt{\frac{\kappa \pm \sqrt{\kappa^2-4\det\mathcal{C}_{AB}}}{2}},
\end{equation} 
and $s_V$ can be expressed
\begin{equation}\label{efficiency3}
s_V(w) = \bigg(\frac{2w+1}{2}\bigg)\log\bigg(\frac{2w+1}{2}\bigg) - \bigg(\frac{2w-1}{2}\bigg)\log\bigg(\frac{2w-1}{2}\bigg),
\end{equation} 
where $\kappa = \det\mathcal{C}_c + \det\mathcal{C}_n + 2\det\mathcal{C}_{cn}$.

\subsection{Results and discussions}

In this section, we will explore how light (photons) and magnetic excitation's (magnons) interact and share quantum correlations and efficiency in a steady and dynamical state, considering various factors. We've selected parameters that are suitable for experimentation~\cite{Tang16}: $\Omega_c/2\pi=10\times 10^6$ Hz, $\Omega_d/2\pi=10\times 10^6$ Hz, $\lambda_d/2\pi=10^2$ Hz, $\lambda_c/2\pi=\lambda_n/2\pi=1\times 10^6$ Hz, $g_{nc}/2\pi = \mathcal{G}_{nd}/2\pi =3.2\times 10^6$ Hz, and at low temperature $T\,{=}\,10\times 10^{-3}$ K. Under these conditions, the coupling between the magnon mode and cavity mode $g_{nc}$ is significantly weaker than the product of the detuning between the magnon and cavity modes and the mechanical resonance frequency, i.e., $g_{nc}^2 \, {\ll}\, |\tilde{ \delta}_n \delta_c | \,\, {\simeq} \, \Omega_d^2$. In this case, we adopt the approximate of the effective magnomechanical coupling as $\mathcal{G}_{nd} \, {\simeq} \sqrt{2} g_{nd} \frac{\Omega}{\Omega_d}$ $\big[$see Eq.~\eqref{nss}$\big]$, where $\mathcal{G}_{nd}/2\pi =3.2\times 10^6$ Hz leading to $|\langle n \rangle| \simeq 1.1 \times 10^7$ for a 250-$\mu$m-diameter YIG sphere, is regarding the drive magnetic field $\mathcal{B}_0 \simeq 3.9 \times 10^{-5}$ T for $g_{nd}/2\pi \simeq 0.2$ Hz and the drive power $\mathcal{P}=8.9\times 10^{-3}$ W. In this order, one can make the Kerr effect negligible because of the realization of the $K|\langle n \rangle|^3 \ll \omega$.

\begin{figure}[!htb]
\minipage{0.5\textwidth}
  \includegraphics[width=\linewidth]{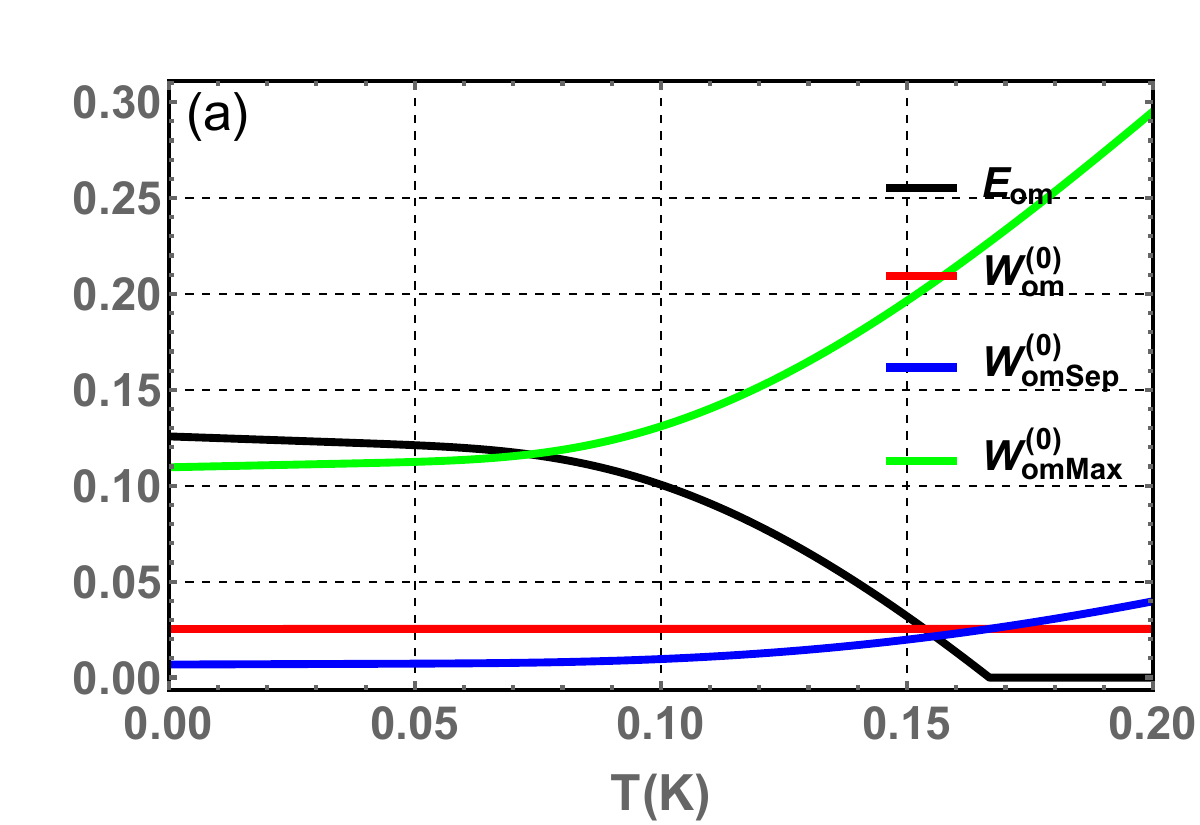}
\endminipage\hfill
\minipage{0.5\textwidth}
  \includegraphics[width=\linewidth]{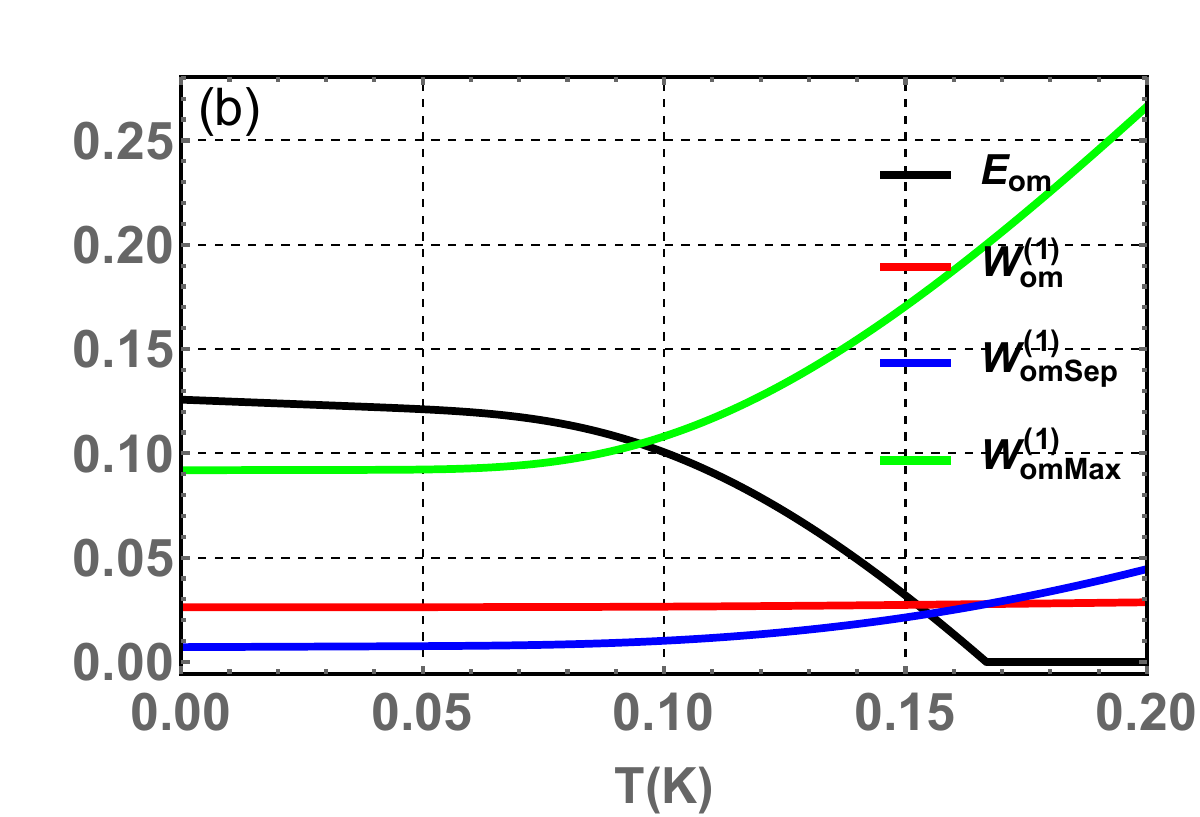}
\endminipage\hfill
\caption{Plot of logarithmic negativity $E_{om}$, extracted work $W_{om}^{(\lambda)}$ (in units of $k_B T$), maximum of extractable work $W_{omMax}^{(\lambda)}$ and extracted work at separable state $W_{omSep}^{(\lambda)}$ between photon and magnon against temperature $T$ for various Gaussian measurements. (a) $\lambda=0$ (homodyne); (b) $\lambda=1$ (heterodyne).}
\label{fig1}
\end{figure}

In Fig. (\ref{fig1}), we plot the logarithmic negativity $E_{om}$, extractable work $W_{om}^{(\lambda)}$ (in units of $k_B T$), separable work $W_{om_\text{sep}}^{(\lambda)}$ and maximum work $W_{om_\text{max}}^{(\lambda)}$ between optical mode and magnon mode versus the temperature $T$ for different measurements. The extractable work $W_{om}^{(\lambda)}$ and separable work $W_{om_\text{sep}}^{(\lambda)}$ are always bound by maximum work $W_{om_\text{max}}^{(\lambda)}$, as depicted in Fig. (\ref{fig1}). We remark that photon and magnon modes are entangled in the region where $W_{om}^{(\lambda)}>W_{om_\text{sep}}^{(\lambda)}$. This agrees with entanglement quantified by logarithmic negativity $E_{om}$~\cite{Jie19}. This figure exhibits that $W_{om}^{(\lambda)}$, $W_{om_\text{sep}}^{(\lambda)}$ and $W_{om_\text{max}}^{(\lambda)}$ all increase with increasing temperature. Conversely, logarithmic negativity diminishes to zero around 0.17 K., i.e., the two modes photon and magnon are in separable state and $W_{om}^{(\lambda)}\leq W_{om_\text{sep}}^{(\lambda)}$, as depicted in Fig. (\ref{fig1})(a-b). We note that for a large value of the temperature $T$ the mode corresponds to the optimal performance of a Szilard engine. This is for homodyne and heterodyne detection ($\lambda=0,1$). Besides, the maximum work is larger at high temperatures. Furthermore, in homodyne detection, the maximum work $W_{om_\text{max}}^{(\lambda)}$ (in units of $k_B T$) achieves 0.30 at $T=0.2$ K (a), while in heterodyne detection it achieves 0.27, as depicted in figure (\ref{fig1}). Thus, one can say that extractable work provides a sufficient condition to witness entanglement in generic two-mode states, which is also necessary for squeezed thermal states.

\begin{figure}[!htb]
\minipage{0.5\textwidth}
  \includegraphics[width=\linewidth]{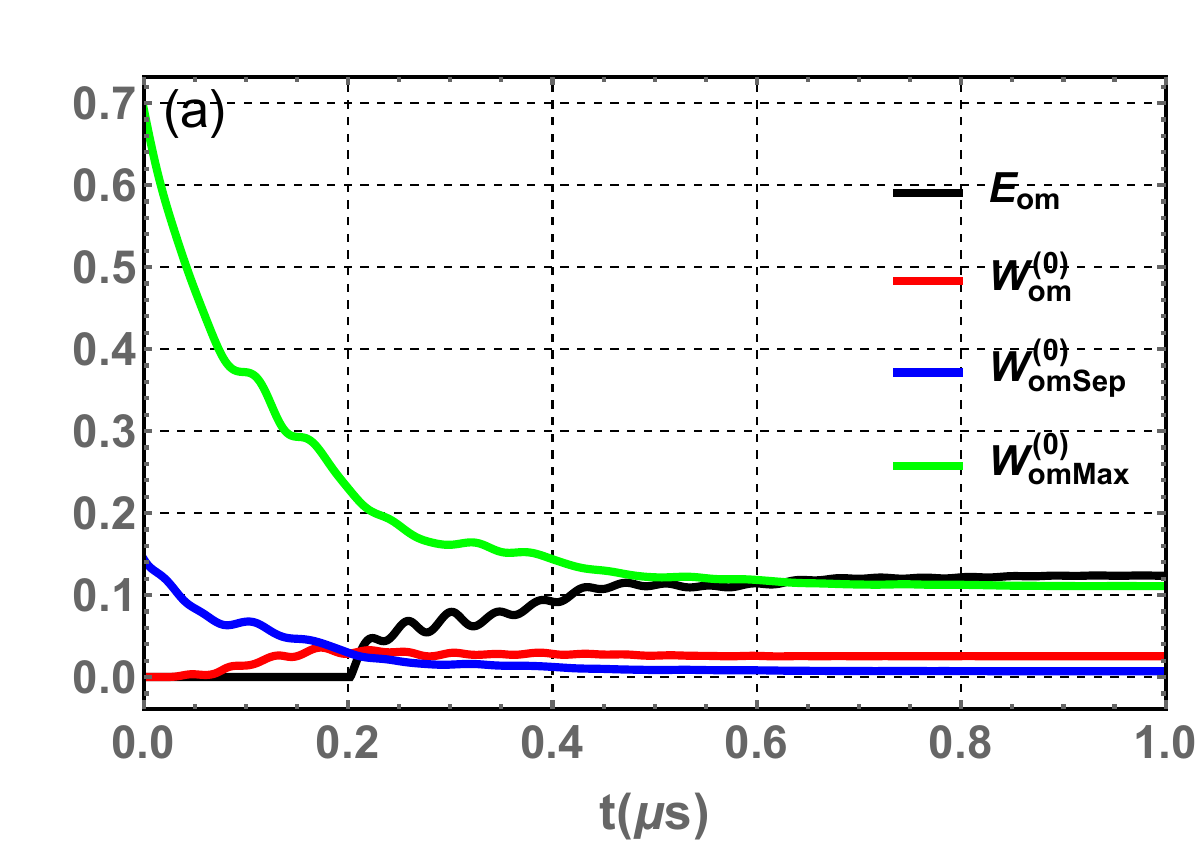}
\endminipage\hfill
\minipage{0.5\textwidth}
  \includegraphics[width=\linewidth]{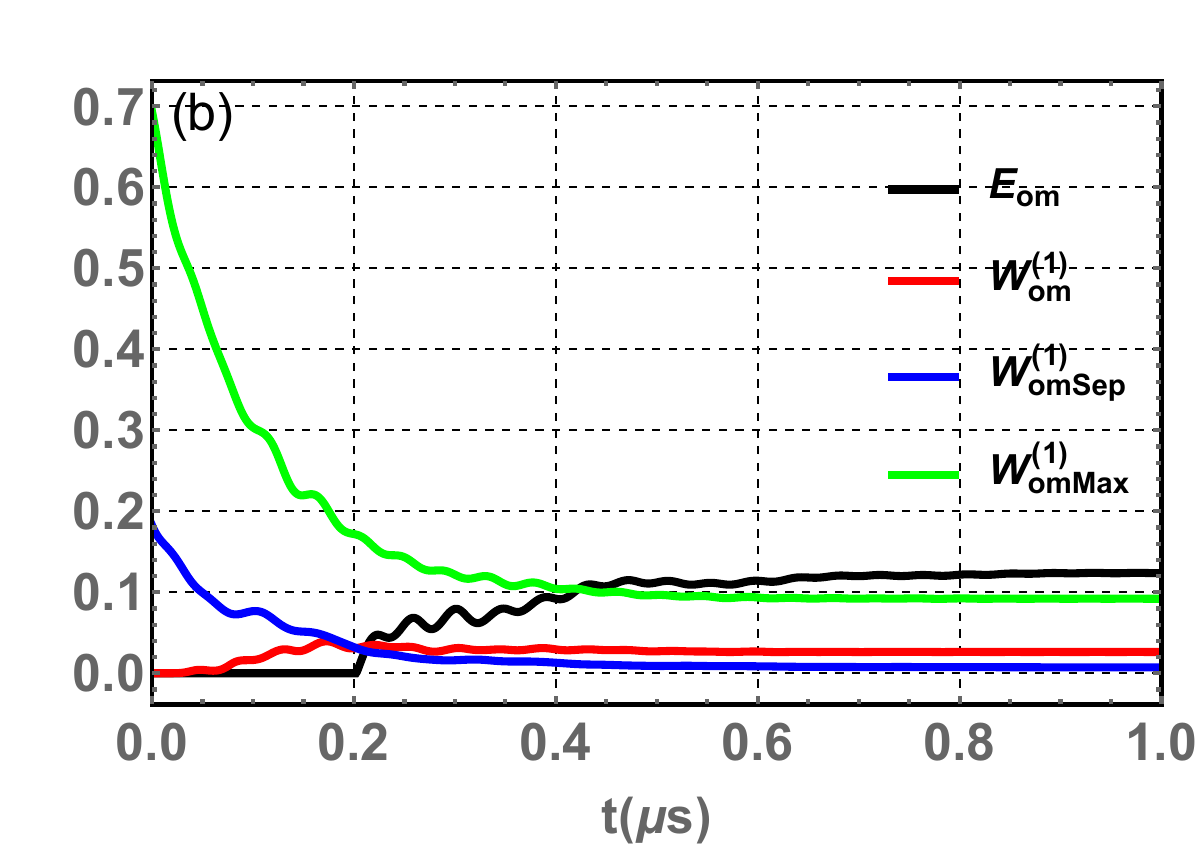}
\endminipage\hfill
\caption{Time evolution of logarithmic negativity $E_{om}$, extracted work $W_{om}^{(\lambda)}$ (in units of $k_B T$), maximum of extractable work $W_{omMax}^{(\lambda)}$ and extracted work at separable state $W_{omSep}^{(\lambda)}$ between photon and magnon for various Gaussian measurements. (a) $\lambda=0$ (homodyne) (b) $\lambda=1$ (heterodyne).}
\label{fig2}
\end{figure}

In Fig. \ref{fig2}, we plot the time-evolution of the bipartite entanglement $E_{om}$, extractable work $W_{om}^{(\lambda)}$ (in units of $k_B T$), separable work $W_{om_\text{sep}}^{(\lambda)}$ and maximum work $W_{om_\text{max}}^{(\lambda)}$ between optical mode and magnon mode in homodyne measurement (a) and in hetrodyne measurement (b). This figure shows three entanglement regimes: The first regime is dedicated to classically correlated states ($E_{om}=0$), i.e., $W_{om}^{(\lambda)}<W_{om_\text{sep}}^{(\lambda)}$. This means that the two modes (photon and magnon) are separated. Nevertheless, the extracted work increases in time while the separable work and maximum work decrease due to decoherence from the thermal bath. The second regime, $W_{om}^{(\lambda)}>W_{om_\text{sep}}^{(\lambda)}$ and $E_{om}>0$, indicating entanglement sudden death between the two modes. Here, we observe the generation of oscillations, which can be explained by the Sorensen-Molmer entanglement dynamics discussed in Ref.~\cite{amaziougsr,Jie17}. The last regime corresponds to a steady state, i.e., $W_{om_\text{sep}}^{(\lambda)}$ remains bounded by $W_{om}^{(\lambda)}$ and $E_{om}$ is constant. The extracted work $W_{om}^{(\lambda)}$ and separable work $W_{om_\text{sep}}^{(\lambda)}$ are bounded by the maximum work $W_{om_\text{max}}^{(\lambda)}$. Thus, the engine has the best performance for strongly squeezed vacuum states and small times of evolution.

\begin{figure}[!htb]
\minipage{0.5\textwidth}
  \includegraphics[width=\linewidth]{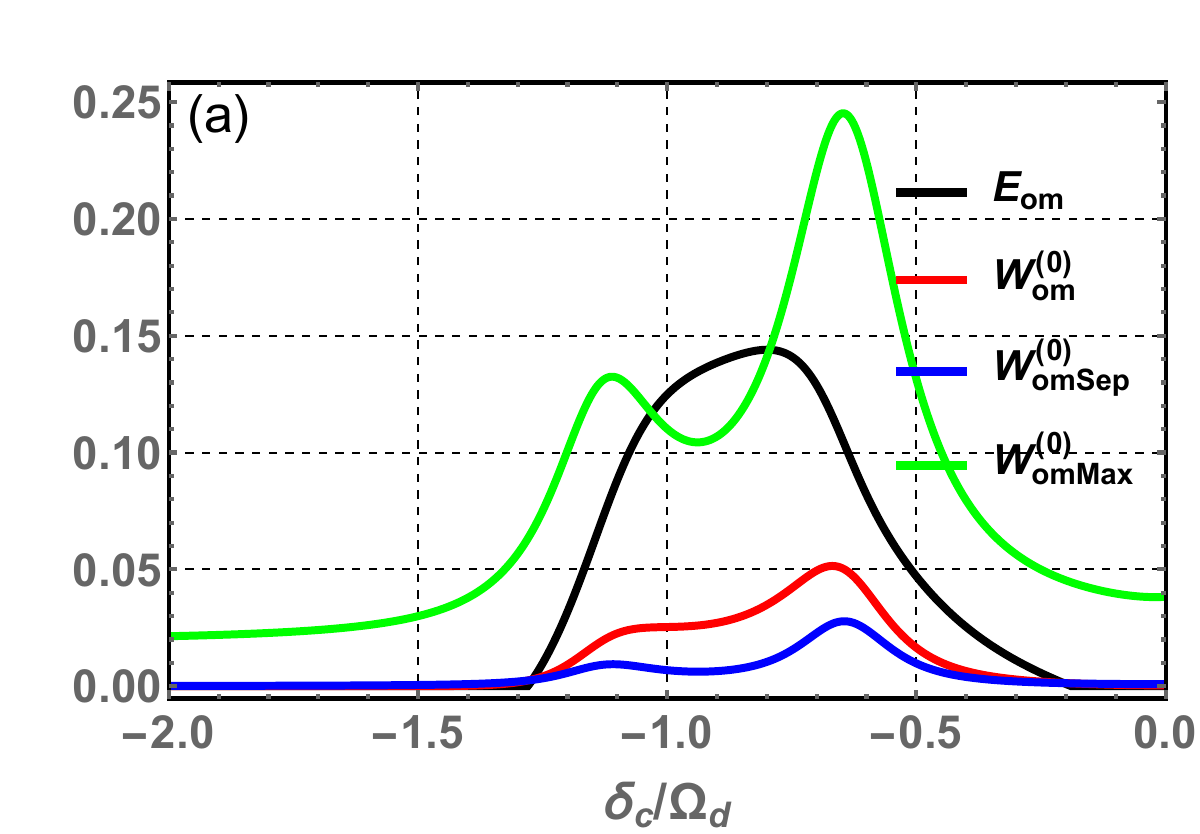}
\endminipage\hfill
\minipage{0.5\textwidth}
  \includegraphics[width=\linewidth]{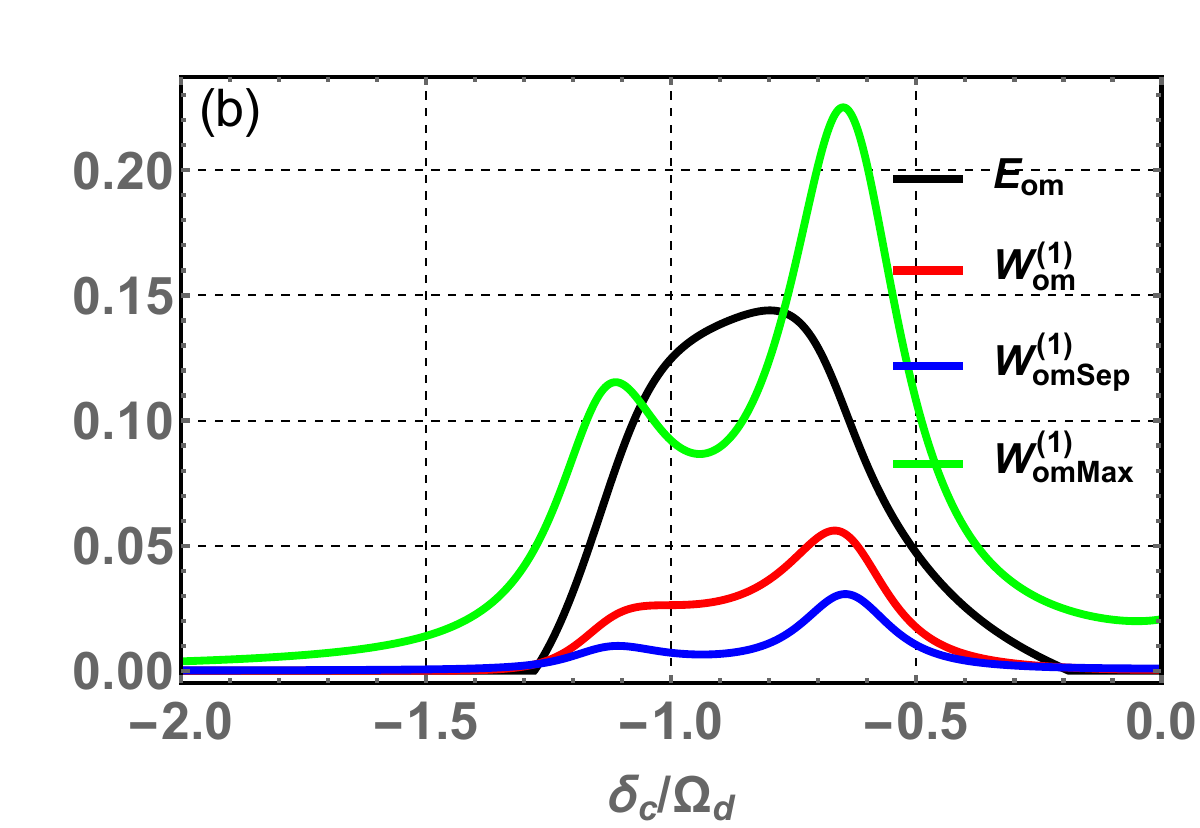}
\endminipage\hfill
\caption{Plot of logarithmic negativity $E_{om}$, extracted work $W_{om}^{(\lambda)}$ (in units of $k_B T$), maximum of extractable work $W_{omMax}^{(\lambda)}$ and extracted work at separable state $W_{omSep}^{(\lambda)}$ between photon and magnon versus the normalized photon detuning for various Gaussian measurements. (a) $\lambda=0$ (homodyne) (b) $\lambda=1$ (heterodyne).}
\label{fig3}
\end{figure}

Figure~\ref{fig3} presents the influence of normalized photon detuning $\delta_c/\omega_d$ on logarithmic negativity $E_{om}$, extractable work $W_{om}^{(\lambda)}$ (in units of $k_B T$), separable work $W_{om_\text{sep}}^{(\lambda)}$, and maximum work $W_{om_\text{max}}^{(\lambda)}$ as functions of normalized magnon detuning $\delta_a/\omega_b$. Entanglement $W_{om}^{(\lambda)}>W_{om_\text{sep}}^{(\lambda)}$, as expected from the logarithmic negativity $E_{om}$~\cite{Jie19}, is observed in Fig.~\ref{fig3}, while the separable state $E_{om} = 0$ and $W_{om}^{(\lambda)} \leq W_{om_\text{sep}}^{(\lambda)}$ is depicted in Fig.~\ref{fig3}(a-b). We remark that the peak in the logarithmic negativity corresponds to the dip in $W_{om}^{(\lambda)}$ and $W_{om_\text{sep}}^{(\lambda)}$ for both homodyne and heterodyne measurements.

\begin{figure}[!htb]
\minipage{0.5\textwidth}
  \includegraphics[width=\linewidth]{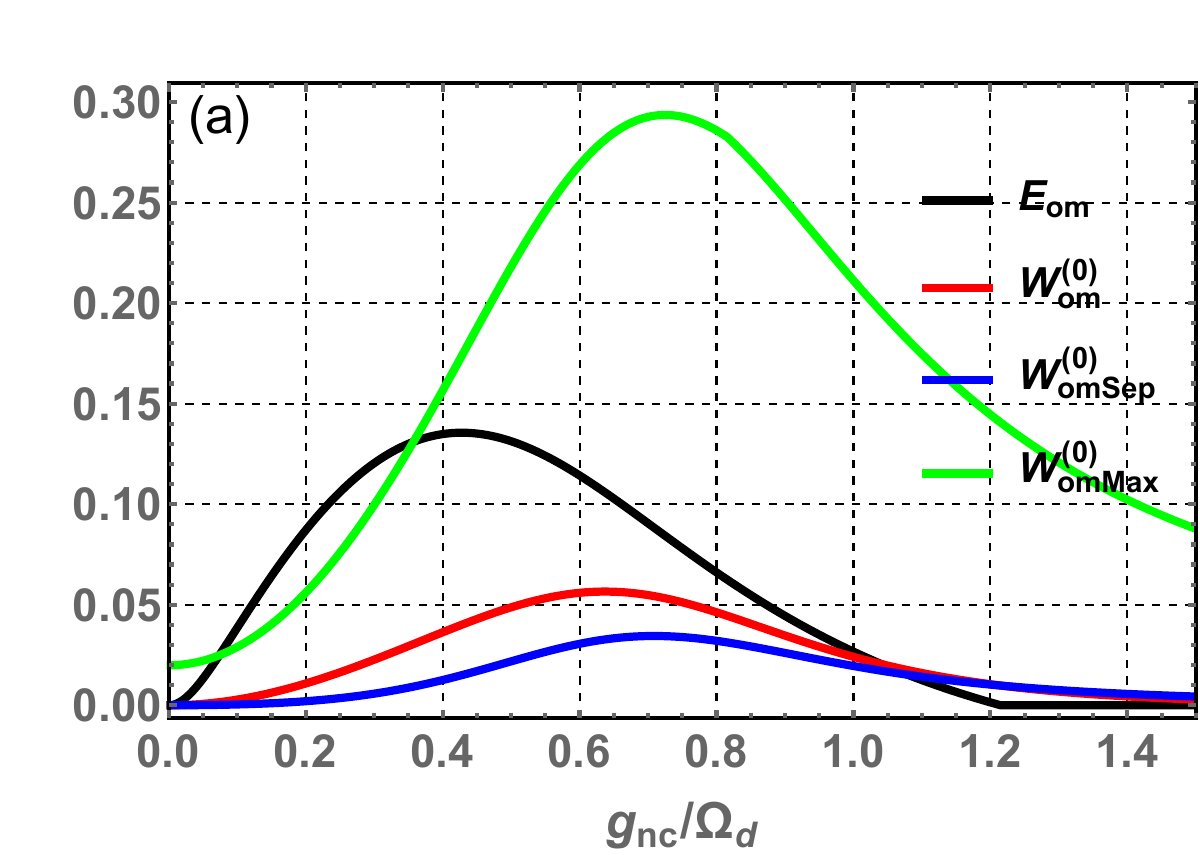}
\endminipage\hfill
\minipage{0.5\textwidth}
  \includegraphics[width=\linewidth]{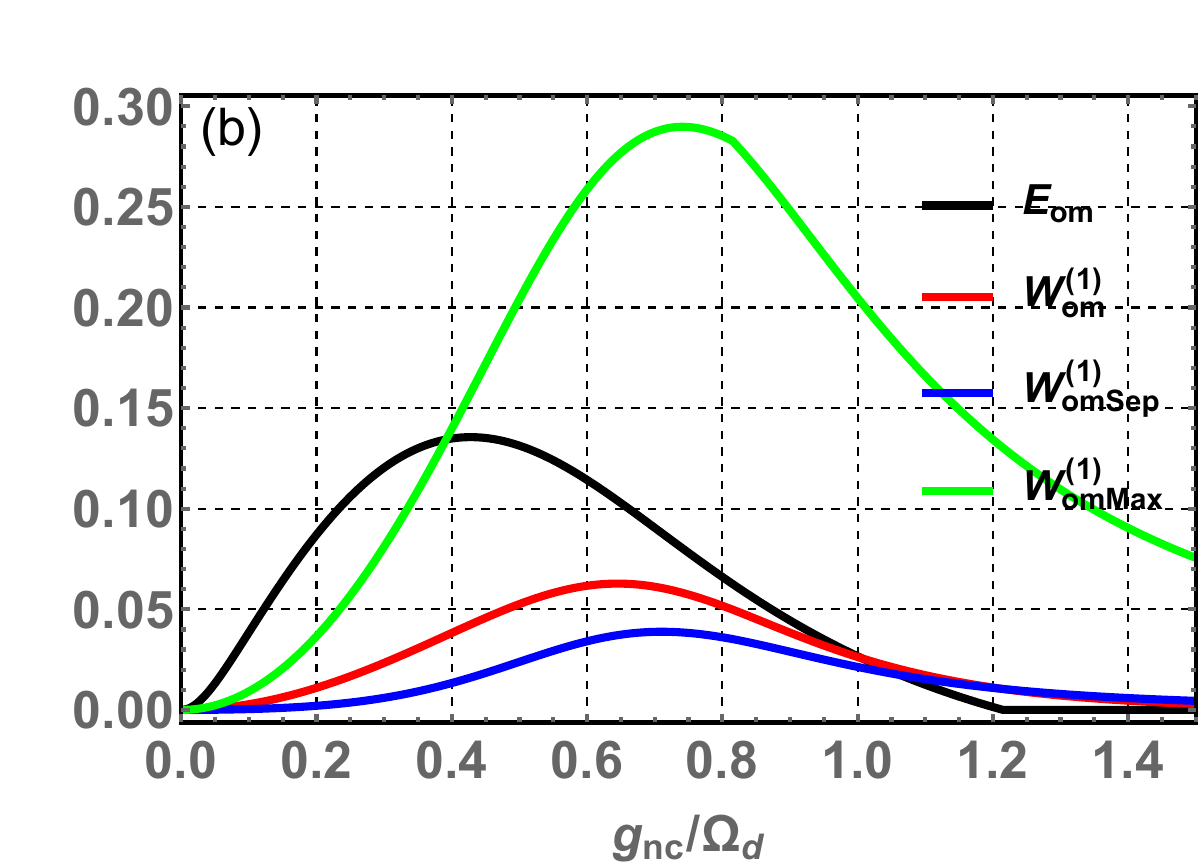}
\endminipage\hfill
\caption{Plot of logarithmic negativity $E_{om}$, extracted work $W_{om}^{(\lambda)}$ (in units of $k_B T$), maximum of extractable work $W_{omMax}^{(\lambda)}$ and extracted work at separable state $W_{omSep}^{(\lambda)}$ between photon and magnon as a function of the magnon-photon coupling $g_{cn}/\omega_d$ for various Gaussian measurements. (a) $\lambda=0$ (homodyne) (b) $\lambda=1$ (heterodyne).}
\label{fig3}
\end{figure}

Figure~\ref{fig4} explores the logarithmic negativity $E_{om}$, extractable work $W_{om}^{(\lambda)}$ (in units of $k_B T$), separable work $W_{om_\text{sep}}^{(\lambda)}$, and maximum work $W_{om_\text{max}}^{(\lambda)}$ between the optical mode and magnon mode versus temperature for both homodyne ($\lambda=0$) and heterodyne ($\lambda=1$) measurements, as a function of the magnon-photon coupling $g_{cn}/\omega_d$. As expected, entanglement $W_{om}^{(\lambda)}>W_{om_\text{sep}}^{(\lambda)}$ coincides with non-zero logarithmic negativity $E_{om}$~\cite{Jie19}. Conversely, the separable state $W_{om}^{(\lambda)}\leq W_{om_\text{sep}}^{(\lambda)}$ and $E_{om}=0$ indicates separability, as shown in Fig.~\ref{fig4}(a-b). Interestingly, Fig.~\ref{fig4} also shows that $E_{om}$, $W_{om}^{(\lambda)}$, $W_{om_\text{sep}}^{(\lambda)}$, and $W_{om_\text{max}}^{(\lambda)}$ all increase with increasing magnon-photon coupling ($g_{cn}/\omega_d$) before gradually decreasing after reaching a maximum value. Additionally, we observe that in homodyne detection, $W_{om_\text{max}}^{(\lambda)}>0$ even for $g_{cn}/\omega_d = 0$, whereas in heterodyne detection, $W_{om_\text{max}}^{(\lambda)}=0$ for $g_{cn}/\omega_d = 0$.

\begin{figure}[!htb]
\minipage{0.6\textwidth}
  \includegraphics[width=\linewidth]{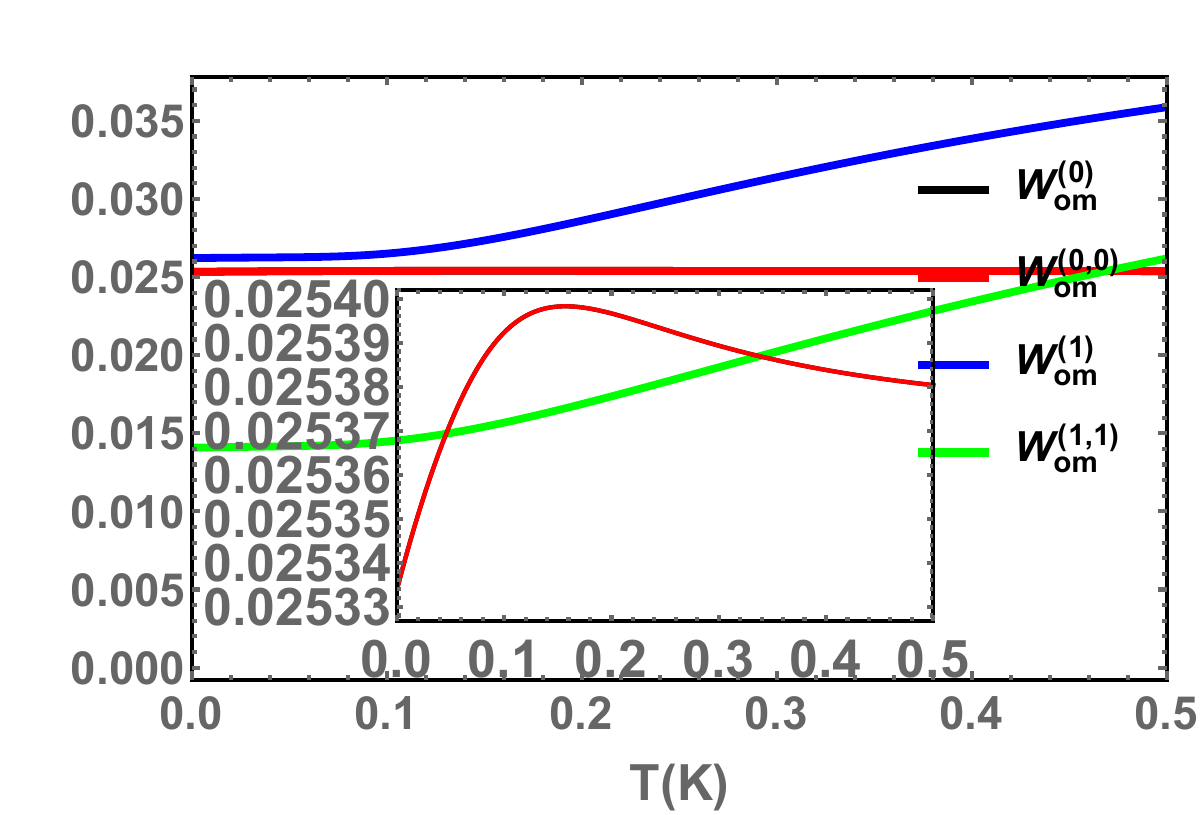}
\endminipage\hfill
\caption{Plot of the extractable work $W$ (in units of $k_B T$) as a function of the temperature $T$ for both measurement $W_{om}^{(0,0)}$ and $W_{om}^{(1,1)}$ and single homodyne measurement $W_{om}^{(0)}$ and heterodyne measurement $W_{om}^{(1)}$.}
\label{twomeas}
\end{figure}

In Fig. \ref{twomeas}, we represent the comparison between the extracted work from both measurement $W_{om}^{(0,0)}(0,0)$ and $W_{om}^{(1,1)}$ to a single homodyne measurement $W_{om}^{(0)}$ and heterodyne measurement $W_{om}^{(1)}$. The decrease in extractable work observed in Fig. \ref{twomeas} is attributed to the second measurement introducing entropy into the system, which can be mathematically represented as a smearing of the distribution imparted by the single measurement, i.e., $W_{om}^{(1,1)}<W_{om}^{(1)}$, just for homodyne detection appears like identical $W_{om}^{(0,0)}<W_{om}^{(0)}$. 

\begin{figure}[!htb]
\minipage{0.5\textwidth}
  \includegraphics[width=\linewidth]{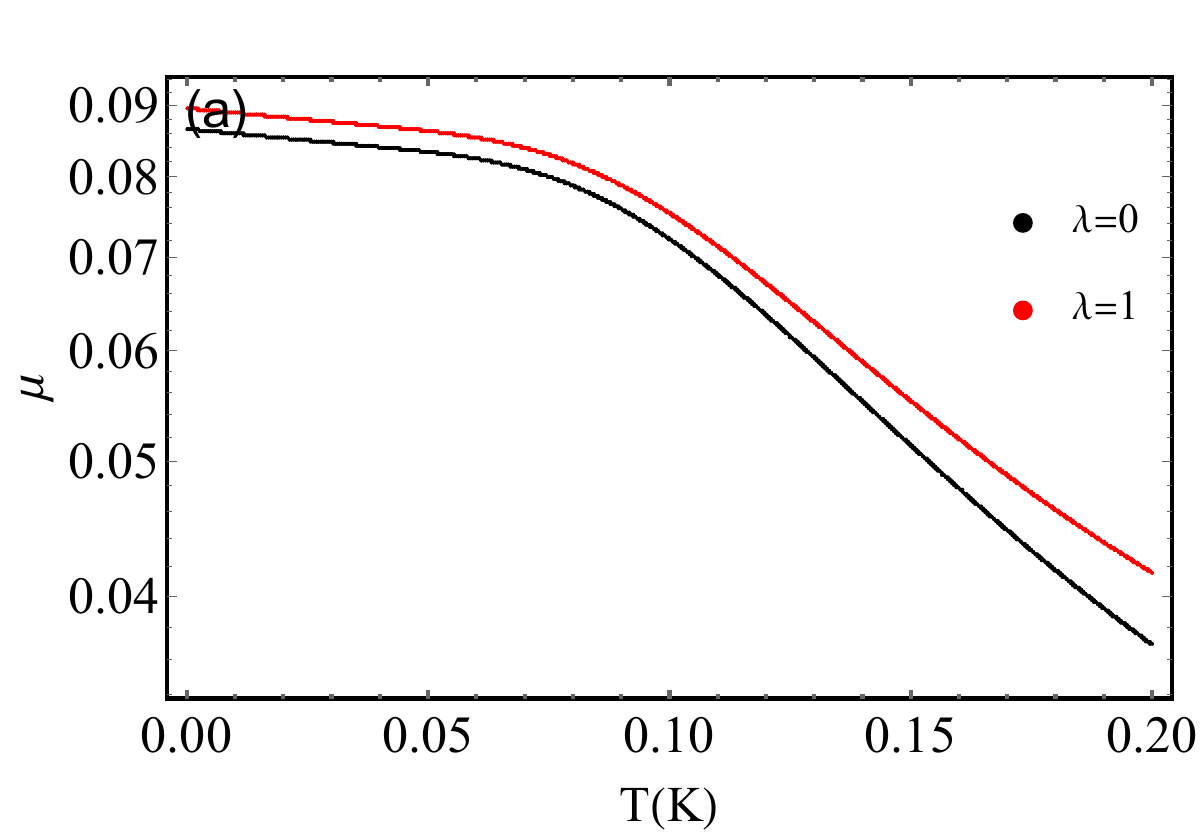}
\endminipage\hfill
\minipage{0.5\textwidth}
  \includegraphics[width=\linewidth]{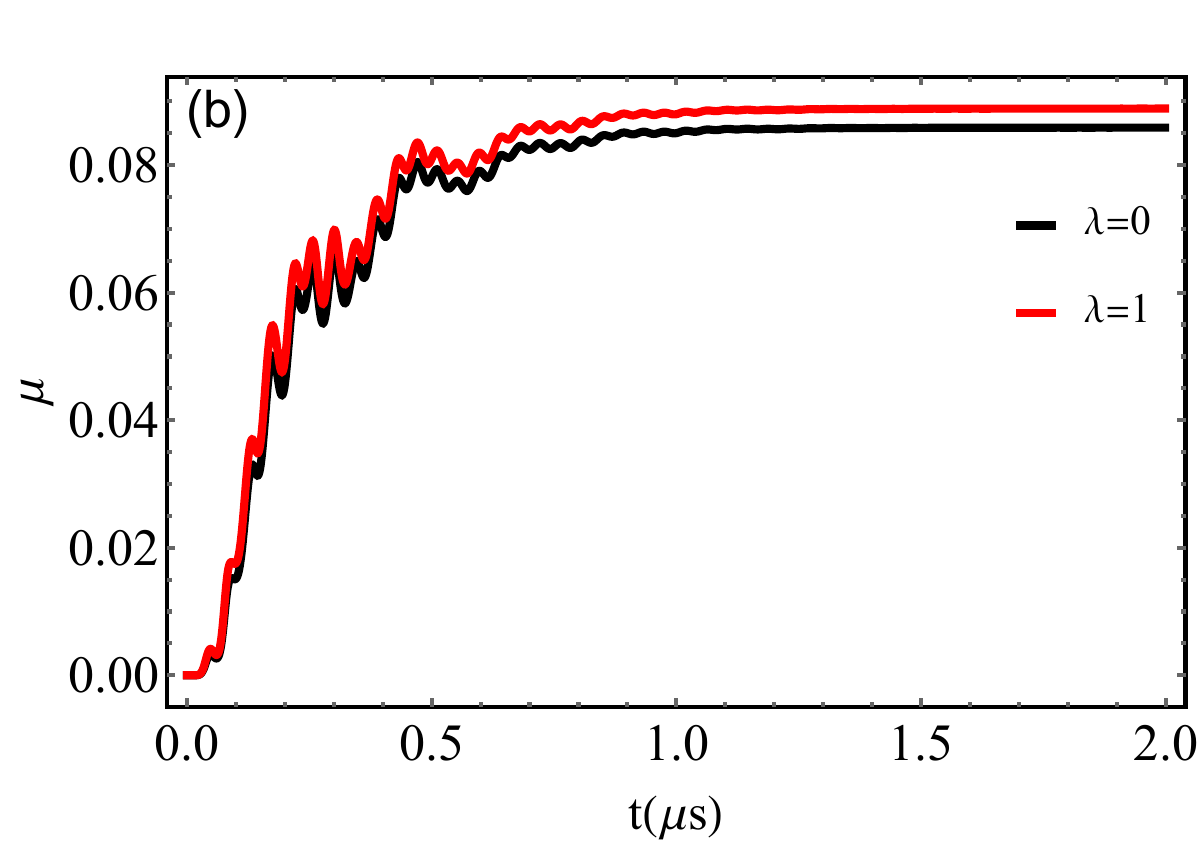}
\endminipage\hfill
\caption{Plot of the efficiency of the work extraction as a function of (a) temperature $T$ and (b) time versus $t(\mu s)$ for various Gaussian measurements with $\lambda=0$ (homodyne) and $\lambda=1$ (heterodyne).}
\label{figeff}
\end{figure}

Fig. \ref{figeff}(a) shows information-work efficiency monotonically decreasing towards zero with increasing temperature for both homodyne and heterodyne detection ($\lambda=0,1$). The engine performs best at low temperatures. Notably, homodyne and heterodyne measurements achieve similar efficiency. Fig. \ref{figeff}(b) explores the time-dependence of efficiency. Here, we see efficiency monotonically increasing to reach a steady-state value, indicating better efficiency for longer times. Besides, the efficiency for homodyne measurement is bound by homodyne measurement, as depicted in Fig. \ref{figeff}.

\section{Conclusion}

In summary, we have discussed a possible strategy to measure the entanglement and separability of the two-mode Gaussian state in a steady and dynamical state by harnessing the extracted work (out of a thermal bath) by means of a correlated quantum system subjected to measurements. Our investigation focuses on a cavity magnomechanical system. Here, a microwave cavity mode is coupled with a magnon mode in a Yttrium Iron Garnet (YIG) sphere. This magnon mode further couples to a mechanical mode through the magnetostrictive interaction. We have used logarithmic negativity to quantify the entanglement between photons and magnons with experimentally reachable parameters. We have shown that the detecting of the entanglement between optic and magon modes is realized when $W_{om}^{(\lambda)}>W_{om_\text{sep}}^{(\lambda)}$ $(\lambda=1,2)$, and it is in agreement with logarithmic negativity. We show that when $W_{om}^{(\lambda)}\leq W_{om_\text{sep}}^{(\lambda)}$ $(\lambda=1,2)$. Besides, extracted and separable work are always bounded by the maximum work in homodyne and hetrodyne measurement. We have found that all works are influenced by different parameters like temperature $T$, normalized magnon detuning $\delta_c/\omega_d$, magnon-photon coupling $g_{nc}$. Also, we have shown that the information-work efficiency of Szilard engine is better for small temperature and for large time. Besides, the efficiency enhances in heterodyne measurement with request to the homodyne measurement.

\end{document}